\documentclass[draftclsnofoot,onecolumn]{IEEEtran}

\usepackage{braket,bbm,amsthm,caption,subcaption}
\usepackage[tbtags]{amsmath} % defines many math commands and
\usepackage{amssymb}  % get, among others, blackboard bold fonts
\usepackage{verbatim} % get comment environment, + new verbatim
\usepackage{amsxtra}  % get, eg, \accentedsymbol

\usepackage{enumerate}

\usepackage{amsfonts}
\usepackage{color}

\usepackage{cite}

\usepackage{mathtools}
\usepackage{url}

\usepackage{epsfig, graphicx, psfrag}

\usepackage[mathcal]{euscript}
\newenvironment{frcseries}{\fontfamily{frc}\selectfont}{}

\usepackage{ifthen}
\providecommand{\VersionLength}{long}
\newcommand{\ver}{\ifthenelse{\equal{\VersionLength}{long}}}
\newcommand{\nver}{\ifthenelse{\equal{\VersionLength}{short}}}

\newtheorem{thm}{Theorem}
\newtheorem{lemma}{Lemma}

\newtheorem{corol}{Corollary}
\newtheorem{prop}{Proposition}
\newtheorem{defn}{Definition}

\newcommand{\Comment}[1]{}
\newcommand{\old}[1]{}
\newcommand{\rem}[1]{}

\providecommand{\comment}[1]{}

\newcommand{\beqn}[1]{\begin{eqnarray}\label{#1}}
\newcommand{\eeqn}{\end{eqnarray}}
\newcommand{\beq}[1]{\begin{equation}\label{#1}}
\newcommand{\eeq}{\end{equation}}

\makeatletter
\newcommand{\vast}{\bBigg@{4}}
\newcommand{\Vast}{\bBigg@{5}}
\makeatother 

\newcommand {\realsplus}{\mathbb{R}_+}

\def \({\left(}
\def \){\right)}
\def \noi{{\noindent}}
\def \noN{\nonumber}

\title{\huge Optimal Discrimination Between Two Pure States \\ and Dolinar-Type Coherent-State Detection} % Article title
\author{Itamar Katz, Alex Samorodnitsky and Yuval Kochman % Leave empty
\thanks{This work was presented in part in ITW 2020, virtual, Apr. 2021, and in the 58th Allerton conference, Monticello, IL, Sep. 2022}
}
\date{}% Leave empty to omit a date
\begin{document}
% Print the title
\maketitle
%--------------------------------------------------------------------
% PRINCIPAL

\begin{abstract}
    We consider the problem of discrimination between two pure quantum states. It is well known that the optimal measurement under both the error-probability and log-loss criteria is a projection, while under an ``erasure-distortion'' criterion it is a three-outcome positive operator-valued measure (POVM). These results were derived separately. We present a unified approach which finds the optimal measurement under any distortion measure that satisfies a convexity relation with respect to the Bhattacharyya distance. Namely, whenever the measure is relatively convex (resp. concave), the measurement is the projection (resp. three-outcome POVM) above. The three above-mentioned results are obtained as special cases of this simple derivation. As for further measures for which our result applies, we prove that Renyi entropies of order $1$ and above (resp. $1/2$ and below) are relatively convex (resp. concave).
A special setting of great practical interest, is the discrimination between two coherent-light waveforms. In a remarkable work by Dolinar it was shown that a simple detector consisting of a photon counter and a feedback-controlled local oscillator obtains the quantum-optimal error probability. Later it was shown that the same detector (with the same local signal) is also optimal in the log-loss sense. By applying a similar convexity approach, we obtain in a unified manner the optimal signal for a variety of criteria. 

\end{abstract}
\section{Introduction}

The problem of optimal discrimination between states is a classical question in the theory of quantum detection and information, see e.g. the survey~\cite{Barnett:09} and the references therein. In this work we consider the simplest case, namely two \emph{pure states}, say $\ket{s_0}$ and $\ket{s_1}$, with priors $(1-\pi,\pi)$. It is well known that the minimum error probability is given by the Helstrom bound \cite{Helstrom1976quantum}:
\begin{align} \label{eq:Helstrom}
p_e \geq \frac{1}{2}\left [1 - \sqrt{1-4\pi(1-\pi) |\braket{s_0|s_1}|^2}\right ].
\end{align}
It is achieved by a projective measurement, where the basis is in the subspace spanned by the states. It was also shown \cite{Levitin94} that the same measurement maximizes the mutual information (MI) between the state and measurement outcome. If, on the other hand, one is interested in unambiguous discrimination, i.e., being able to reach certainty regarding the state with high probability, the optimal strategy is a three-element von Neumann measurement; two of the three outcomes lead to certainty (one for each state) and the third is uncertain \cite{Unambiguous2,Unambiguous3,Unambiguous}. 

These optimal solutions have nice symmetry properties, in terms of the \emph{backward} channel from the estimated state to the true state, even if the prior $\pi$ is not uniform. For the projection, the error probability is independent of the outcome, and for the three-element measurement, the posterior given the uncertain outcome is uniform (for any prior $\pi$ that is not ``too skewed''). These backward channels, which we dub ``the extremal channels'' for reasons that will become clear later, are demonstrated in Figure~\ref{fig:channels}. Although these properties may be known to many, we show them in Appendix~\ref{app:backward} for completeness.

    \begin{figure}[htbp]
        \centering
        \subfloat[\centering Binary channel for minimum error probability]{{\includegraphics[width=6cm]{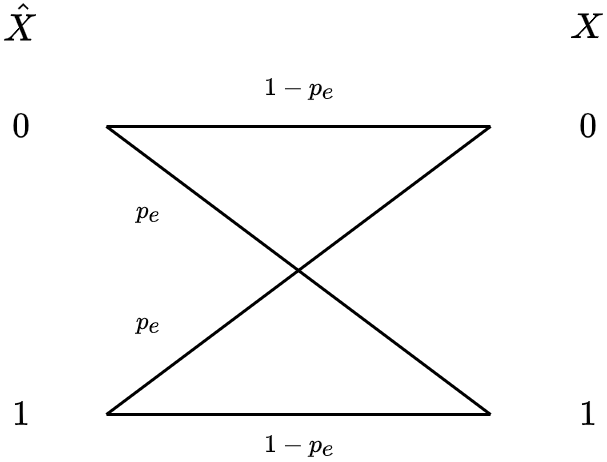}}\label{fig:channel1}} %
        \qquad
        \qquad
        \qquad
        \qquad
        \subfloat[\centering Ternary channel for unambiguous discrimination]{{\includegraphics[width=5.95cm]{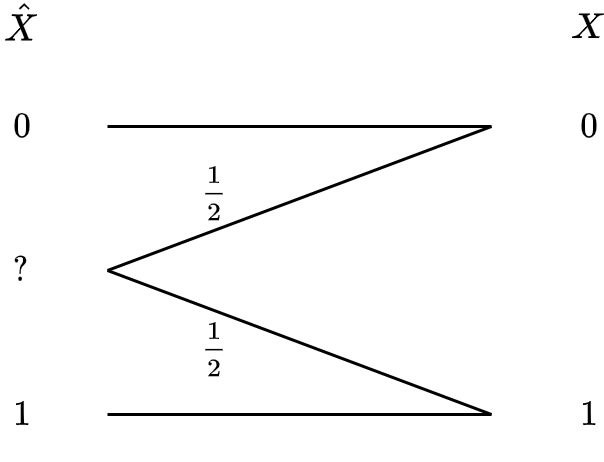}}\label{fig:channel2}}%
        \caption{Extremal backward channels.}\label{fig:channels}
    \end{figure}

The results for minimum error probability, maximum MI and unambiguous discrimination were derived and proven separately. In this work we present a framework which allows to prove all of them in a unified manner that is very simple, and furthermore to obtain optimality results for other criteria that satisfy certain convexity conditions. Namely, denote by $(p,1-p)$ the posterior distribution of the states, which is itself a random variable which depends on the outcome. Then the three problems can be described as finding the measurement which obtains 
\begin{align} \label{eq:objective} \min E[g(p)], \end{align}
where $g$ is some \emph{objective function} and the expectation is taken over the state prior and the measurement outcome. The objectives for the three problems are the error-probability function, the binary entropy and the ambiguity function, given by
\begin{subequations}
\begin{align}
    e(p) &= \min(p,1-p) \label{eq:error} \\
    h(p) &= p \log\frac{1}{p} + (1-p)\log\frac{1}{1-p} \label{eq:entropy} \\
    a(p) &= \mathbbm{1}_{p\notin\{0,1\}}, \label{eq:ambiguity}
\end{align}
\end{subequations}
respectively. 

We consider objective functions that satisfy a convexity condition w.r.t. the Bhattacharyya parameter
\begin{align} \label{eq:Bhattacharyya}
    b(p) = \sqrt{p(1-p)}.
\end{align}
As we discuss in the sequel, we can have a partial ordering of functions in terms of convexity: for functions $g_1(\cdot)$ and $g_2(\cdot)$ we say that $g_1$ is convex with respect to $g_2$, denoted as $g_1 \succ g_2$, if there exists a convex $h(\cdot)$ such that $g_1(\cdot)=h(g_2(\cdot))$. We then say that an objective function $g$ is convex- (resp. concave-) admissible if it is symmetric around $1/2$ and $g \succ b $ (resp. $b \succ g$). The class of convex-admissible objectives includes $e(p)$ and $h(p)$ above, as well as the min entropy; the class of concave-admissible functions includes $a(p)$ above.  We show that all R\'enyi entropies of high enough or low enough order, also fall into one of the classes. 

Our main result states, that for any convex-admissible objective the optimal measurement is the symmetric projection optimal for error probability, while for any concave-admissible objective (as long as the prior is ``not too skewed'' in a sense defined in the sequel) the optimal measurement is the three-element von Neumann measurement optimal for unambiguous discrimination; further, in both cases the optimal performance satisfies the simple relation\footnote{Even though we minimize $E[g(p)]$, since $g$ is a function of $b$ we may obtain such an expression. For example one may verify that it agrees with the Helstrom bound on the error probability \eqref{eq:Helstrom} when $b(p)$ is deterministic.}
\begin{align} 
E[ b(p) ]  = \braket{s_0|s_1} b(\pi). \label{eq:root_result}
\end{align}
Our proof technique is extremely simple. First we show that for the Bhattacharyya objective $b$, there are many optimal strategies yielding \eqref{eq:root_result}. These include the extremal channels of Figure~\ref{fig:channels}, which are special in the following sense: under the first $b(p)$ is deterministic, while under the second it obtains either its minimal value $b(0) = 0$ or its maximal value $ b(1/2) = 1/2$. From here the result follows directly: For $g \succ b$, a deterministic $b$ is optimal for $g$ by Jensen's inequality, while for $b \succ g$ a maximally-spread $b$ is optimal for $g$ by an ``inverse Jensen'' lemma.

Even when the optimal quantum measurement can be derived, it is not always clear how to materialize this measurement using a physical system. One case where an optimal measurement scheme was proposed, is detection of coherent optical states. With some abuse of notation, suppose that the states are given by pulses with complex amplitudes $\{s_i(t)\}$, $0 \leq t \leq T$ and $i \in \{0,1\}$, then we have two pure states with
\begin{align} \label{eq:coherent_inner}
 |\braket{s_0,s_1}|^2 = \exp\left\{ -\int_0^T |s_1(t) - s_0(t)|^2 dt \right\} . \end{align}
Dolinar \cite{Dolinar1973} showed that the corresponding Helstrom bound can be obtained by a scheme comprising of a local oscillator producing a coherent state, a beam splitter that combines the received and local signals, and a photon counter, where the local signal depends upon past photon counts. Similar to the projective measurement for general pure states, the error probability of Dolinar's receiver is independent of the measurement outcomes (photon counts). This independence yields favorable properties: the optimal strategy at each time $t$ is independent of the pulse duration $T$, and the local signal is one of two prescribed waveforms, where switching occurs with each photon count. It was shown in \cite{Lizhong2} that Dolinar's scheme and choice of local signal are also optimal in the MI sense. 

Dolinar's receiver has been subject of many works, offering interpretations, implementations and extensions; we survey some of them in the sequel. 
Our contribution is providing an interpretation of the receiver optimality, by applying to the coherent problem the same approach that we use for general coherent states. We find that a continuum of strategies is optimal for the Bhattacharyya function $b$. These include Dolinar's strategy which yields a deterministic $b$, and a strategy which ``cancels'' the more probable signal which yields (under a ``not too skewed'' condition) either minimal or maximal $b$. Using the same convexity arguments as above we have the optimal strategy for relatively convex or relatively concave objectives, and for all of these the performance coincides with the best quantum measurement. It should be noted that the derivation finds the optimal strategy within the Dolinar structure, without resorting to any quantum considerations, beyond the fact that the photon counts constitute a Poisson process with rate proportional to the squared input amplitude.

Finally, we comment about minimax optimality. It can be readily shown that the minimax rule (with respect to the two states) is the Bayesian rule assuming a uniform prior. As with a uniform prior the ``not too skewed'' condition is always satisfied, our analysis immediately gives the minimax rule with respect to any objective function that is convex or concave relative to the Bhattacharyya objective $b$.

\section{Mathematical Background}

\subsection{Convexity: Basic Inequalities}

Let $\phi(\cdot)$ be a convex function, then the well-known Jensen equality states that for any real-valued integrable random variable $X$,
\begin{align} \label{eq:Jensen}
E[\phi(X)] \geq \phi(E[X]),
\end{align}
with equality if and only if $\phi(X)$ equals some affine function of $X$ with probability (w.p.) $1$ (which happens, e.g., if $X$ is deterministic).
If in addition $X$ is bounded, we can have an ``inverse'' bound as follows.
\begin{prop}
Let $X$ and $Y$ be real-valued integrable random variables s.t. $E[X]=E[Y]$. Assume that for some finite $a \leq b$,
\[ \Pr (X\in[a,b]) = \Pr (Y \in\{a,b\}) = 1. \]
Then for any convex $\phi(\cdot)$,
\[ E[\phi(X)] \leq E[\phi(Y)] . \] Equality holds if and only if $\phi(X)=\ell(X)$ w.p. $1$, where $\ell(X)$ is the affine function satisfying $\ell(a)=\phi(a)$ and $\ell(b)=\phi(b)$.
\end{prop}
\begin{proof}
In terms of $\ell(\cdot)$ defined above,
\begin{align*} E[\phi(X)] &\leq E[\ell(X)]
% \\ &= \ell(E[X]) \\ &= \ell(E[Y])
\\ &= E[\ell(Y)] \\ &= E[\phi(Y)], 
\end{align*}
where the inequality holds since by convexity $\phi(X) \leq \ell(X)$ (for $a\leq X \leq b$ which happens w.p. 1), the first equality holds by the assumption of equal means and the linearity of the mean, and the second holds since $\phi(Y) = \ell(Y)$ ($Y$-almost everywhere).
\end{proof}
Notice that the equality condition holds if $X \in \{a,b\}$ w.p. $1$. Also notice that since the distribution of $Y$ can be calculated in terms of $a$, $b$ and the mean, we can re-write this result in terms of the distribution of $X$ only:
\begin{align} \label{eq:inverse_Jensen}
(b-a) E[\phi(X)] \leq (E[X] - a) \phi(b) + (b-E[X]) \phi(a). \end{align}

\subsection{Relative convexity}

For univariate differentiable functions, convexity is a relation of a function with its tangents, i.e., linear functions. Relative convexity defines a relation that is generalized beyond linear functions. For two functions $f$ and $g$ that are defined on the same (open or closed) interval $\mathcal{I}$, we say that $f$ is convex relative to $g$, denoted $f(\cdot)~\succ~g(\cdot)$, if there exists a function $\phi$ that is convex and non-decreasing on the range of $g$ such that $f = \phi(g)$. We say that $f$ is concave relative to $g$, denoted $f(\cdot)~\prec~g(\cdot)$, if there exists such concave $\phi$. This concept goes back to the 1930s \cite{Jessen_Relative}. In ~\cite{palmer2003relative}, Palmer adds the requirement of invertibility of $\phi(\cdot)$, making relative convexity a partial ordering. We avoid this assumption since it doesn't hold in a case that is of high interest, yet we mostly follow Palmer's exposition. 
An equivalent definition of relative convexity from a geometric point of view \cite[Theorem 2]{palmer2003relative} is that 
$f (\cdot) \succ g(\cdot)$ if and only if $\forall x_0 \in \mathcal{I}, \ \exists \lambda \in [0, \infty]$ s.t.
        \begin{IEEEeqnarray}{r.C.l?l}
        	f(x)-f(x_0) &\geq& \lambda(g(x)-g(x_0)) \ \forall x \in \mathcal{I}. \label{eq:relative_convexity_def2}
        \end{IEEEeqnarray}
If $f$ and $g$ are twice differentiable, the following criterion may be simpler to check: $f (\cdot) \succ g(\cdot)$ if and only if \cite[Theorem 4]{palmer2003relative}
\begin{IEEEeqnarray}{r?C?l} 
	 \frac{f''(x)}{|f'(x)|} \geq \frac{g''(x)}{|g'(x)|}  \ \forall x\in \mathcal{I}. \label{eq:derivatives_ratio}
\end{IEEEeqnarray}
Of course, the opposite conditions hold for relative concavity.

\subsection{Admissible Objective Functions}

Recall that we are interested in minimizing the expected value of functions of a binary distribution \eqref{eq:objective}. The class of relevant functions is as follows.
\begin{defn}
Consider a function $g(p)$: $(0,1)\rightarrow\realsplus$.\footnote{Here and in the sequel, $\realsplus$ denotes the non-negative reals.} We say that $g(p)$ is \emph{admissible} if it satisfies:
    \begin{itemize}
    \item monotonicity: $g(p)$ is non-decreasing in $[0,1/2]$.
        \item Symmetry: $g(p)=g(1-p)$.
        \item Normalization: $g(0) = 0$.
        \item Boundedness: $g(1/2)$ is finite.
    \end{itemize} \label{def:admissible}
    \end{defn}
Further, it turns out that convexity relative to the Bhattacharyya objective \eqref{eq:Bhattacharyya} is a key property.
\begin{defn} An admissible objective $g(\cdot)$ is called convex-admissible (resp. concave-admissible) if it satisfies $g \succ b$ (resp. $g \prec b$).
\end{defn}
   
   We can relate relative convexity to the convexity inequalities \eqref{eq:Jensen}-\eqref{eq:inverse_Jensen} by applying them to the convex (or concave) mapping $\phi$ between the functions. Although this can easily be presented for any two functions with a relative convexity relation, for conciseness we only present it for convex- and concave-admissible objectives.
   
   \begin{prop} 
  \begin{itemize}
\item []
       \item 
   Let $g$ be convex-admissible, then
        \begin{align} \label{eq:admissible_Jensen} E[g(p)] \geq g(p^*) \end{align}
        where $p^*$ satisfies $b(p^*) = E[b(p)] $. Equality holds if and only if there exists $p \geq 0$ s.t. $g(p) = b(p) $ w.p. $1$.
  
   \item
   Let $g$ be concave-admissible, then
  
         \begin{align} \label{eq:admissible_inverse_Jensen} E[g(p)] \geq  2 g\left(\frac{1}{2}\right) E[b(p)].  \end{align} Equality holds if and only if 
        \[ g(p) = 2 g\left(\frac{1}{2}\right) b(p) \] w.p. $1$.

   \end{itemize}
   \label{prop:Jensen_admissible}
   \end{prop}
    
    We conclude by noting that a sufficient condition for equality in \eqref{eq:admissible_Jensen} is that $b(p)$ is deterministic, i.e., $p \in \{p_e, 1-p_e\}$ for some value $p_e$; a sufficient condition for equality in \eqref{eq:admissible_inverse_Jensen} is that $b(p)$ is either minimal or maximal, i.e., $p \in \{0,1/2,1\}$. If $p$ is taken to be the probability of some binary variable $X$ given some $\hat X$, then these two cases correspond exactly to the two extremal channels of Figure~\ref{fig:channels}.

\subsection{Examples}

We now consider the relative convexity of some admissible objective functions. 
Throughout we use the fact (that is not hard to show), that verifying relative convexity in $(0,1/2)$ is enough in order to show that an admissible function that is twice-differentiable in $(0,1/2)$ and continuous at 1/2 is a convex-admissible function.

    The Error-Probability function $e(p)$ \eqref{eq:error}. It is convex-admissible since it is linear in $(0,1/2)$ while $b(p)$ is concave.
   
   Now, consider the family of binary Renyi entropies:
      \begin{align} \label{eq:Renyi} h_\alpha(p) = \left\{ \begin{array}{cc}
           \frac{1}{1-\alpha} \log \left( p^\alpha + (1-p)^\alpha \right) & \alpha \in (0,1) \cap (1,\infty) \\
           a(p) & \alpha = 0 \\
           h(p) & \alpha = 1 \\
           -\log(1-e(p)) & \alpha = \infty
       \end{array} \right. . \end{align}
    
    In Appendix~\ref{app:Renyi} we prove the following result, which may be useful also outside the context of this paper.
    
    \begin{thm} \label{thm:Renyi}
    
        Let $\alpha \geq \beta \geq 0$, then $h_\beta(\cdot) \prec  h_\alpha(\cdot)$.

\end{thm}

    Notice that the reverse relation $h_\alpha(\cdot) \succ h_\beta(\cdot)$ does not hold for $\beta=0$. In order to relate this result to the Bhattacharyya function, consider the following two special cases:
    
    \begin{itemize}
   
        \item Binary entropy $h_1(p) = h(p)$ \eqref{eq:entropy}. It is also convex-admissible since the condition \eqref{eq:derivatives_ratio} applied to $b(p)$ amounts to
        \[ \log \frac{1-p}{p} \geq 2 (1-2p), \]
which can be easily verified to hold for all $p\in(0,1/2)$.
        \item $h_{1/2}(p)$ is concave-admissible since 
        \[ h_{1/2}(p) = \log(1+2b(p)) .\]
        \end{itemize}
        
        Thus, we immediately have the following.
        \begin{corol} \label{corol:Renyi}
        \begin{itemize}
            \item For all $\alpha \geq 1$, $h_\alpha(\cdot)$ is convex-admissible.
            \item For all $\alpha \leq 1/2$, $h_\alpha(\cdot)$ is concave-admissible.
        \end{itemize}
        \end{corol}
        
        Indeed, for intermediate values $1/2 < \alpha < 1$, neither convexity nor concavity hold.

\section{Pure States}
\label{sec:pure}

Let $X$ be a Bernoulli variable, and let $\ket{s_0}$ and $\ket{s_1}$ be two states in some complex Hilbert space. We measure a state $\ket{s}$ which is $\ket{s_i}$ if $X=i$, with prior 
\[ \Pr(X=1) = \pi, \]
where w.l.o.g $\pi \leq 1/2$.
Let
$\{E_j\}, j=0,\ldots,J-1$ be a POVM in this space, i.e., the operators are positive semi-definite and
\[ \sum_{j=0}^{j-1} E_j = I. \]
Recall that the probability of outcome j given $X=i$ is given by \[q_{i,j}=\braket{s_i|E_j|s_i}, \]
and denote the outcome probabilities by
\[q_j = (1-\pi) q_{0,j} + \pi q_{1,j}. \] Let $p_j = \Pr(X=1|\text{outcome}=j)$ be the posterior Bernoulli parameter associated with outcome $j$, given by Bayes' rule:
\begin{align} \label{eq:Bayes} p_j = \frac{\pi q_{1,j}}{q_j}, \ j=0,\ldots,J-1 : q_j > 0. \end{align}

\begin{figure}[htbp]
	\centering
	\includegraphics[width=15cm]{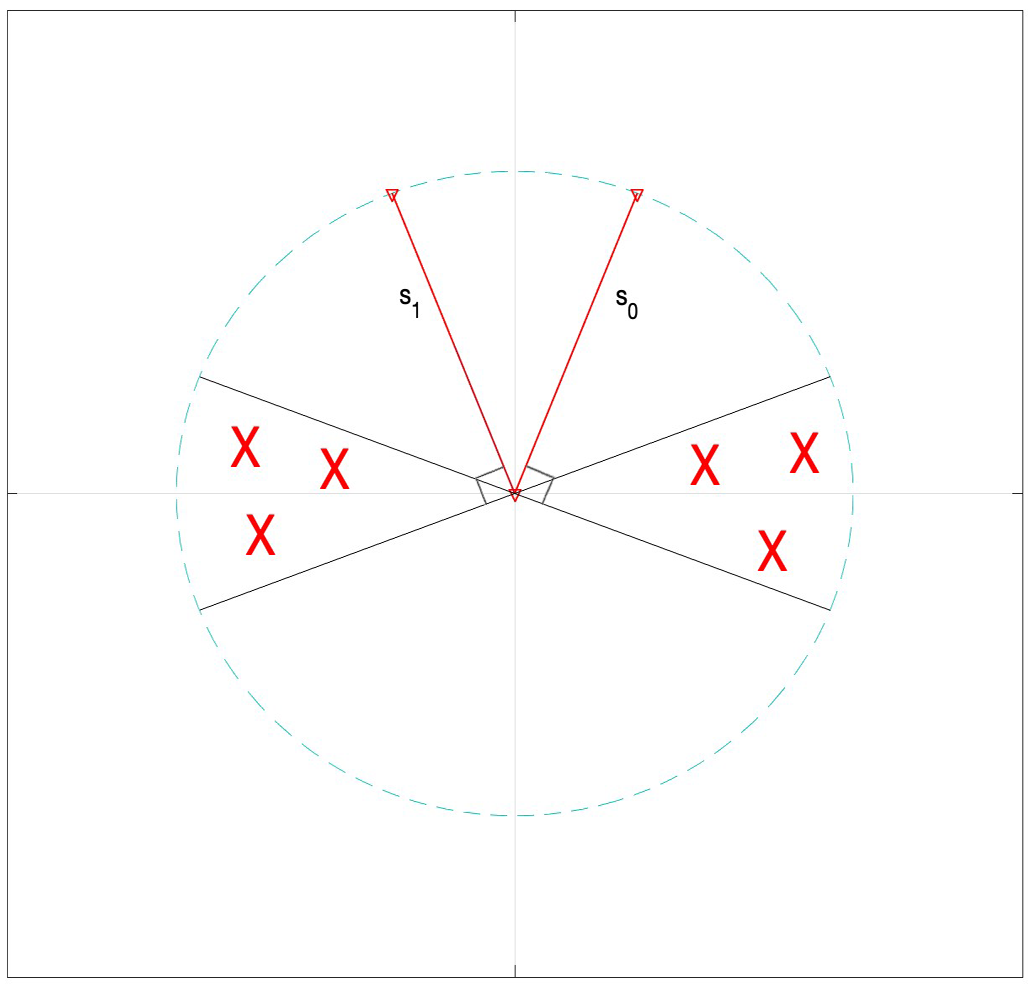}
	\caption{Allowed measurement vectors. The 'X' denote the forbidden angles.}
	\label{fig:allowed}
\end{figure}

The following class of measurement will turn out important.
\begin{defn}
An \emph{eligible measurement} for $\ket{s_0}$, $\ket{s_1}$ is a POVM satisfying, for all $j\in\{0,\ldots,J-1\}$
\begin{enumerate}
    \item $\braket{s_0|E_j|s_0} \braket{s_1|E_j|s_1} = |\braket{s_0|E_j|s_1}|^2$.
    \item $\braket{s_0|E_j|s_1} \in \realsplus$. 
\end{enumerate}
\end{defn}
We interpret this class as follows. Notice that the first condition is that equality holds in Cauchy-Schwartz; a sufficient condition (necessary except for trivial cases) is that $E_j$ is rank-1. We can characterize such measurement by a set of vectors $\{\ket{v_j}\}$ and weights $\{\alpha_j\}$ s.t.
\[ E_j = \alpha_j \ket{v_j}\bra{v_j}, \ j=0,\ldots,J-1. \]
We may ignore global phase and limit our attention to real states, where
the second condition for eligibility becomes
\[ \text{sign} (\braket{s_0|v_j}) = \text{sign} (\braket{s_1|v_j}) \ \forall j=0,\ldots,J-1. \]
Thus, the condition is that for all $j$, either both angles $(s_0,v_j)$, $(s_1,v_j)$ are sharp, or they are both obtuse. We may further limit our attention to the two-dimensional space spanned by the states, where we have the geometric picture depicted in Figure~\ref{fig:allowed}: We may choose any number $J\geq 2$ of vectors with weights chosen to satisfy the POVM condition, as long as none have ``forbidden'' angles with the states. For example, for projections, it means that we cannot choose a vector between the states (because then the second vector, normal to it, will have a forbidden angle).

Now, we say that a problem defined by $(\ket{s_0},\ket{s_1},\pi)$ is \emph{not too skewed} if
\[ \frac{\pi}{1-\pi} \geq |\braket{s_0|s_1}|^2. \]
The following shows the existence of eligible measurements with favorable properties.
\begin{prop} \label{prop:exists_pure}
For any two states $\ket{s_0}$ and $\ket{s_1}$ and any prior $\pi$,
\begin{enumerate}
    \item There exists an eligible measurement with $J=2$, s.t. $p_0+p_1=1$.
    \item If the problem is not too skewed then there exists an eligible measurement with $J=3$, s.t. $p_0=0$, $p_1=1$ and $p_2=1/2$.
\end{enumerate}
\end{prop}
The proof is given in Appendix~\ref{app:backward}. These two kinds of eligible measurements are nothing but the well-known solutions to the minimum error probability and unambiguous discrimination problems, respectively. Notice that the posterior probabilities match exactly the extremal channels of Figure~\ref{fig:channels}.

We are now ready to state our main result. We are given some objective function $g(p): [0,1] \rightarrow \mathbb{R}_+$, where without loss of generality $h(0)=0$. For any POVM, the expected objective is given by
\[ \bar g = \sum_{j=0}^{J-1} q_j g(p_j). \] 
Let $\bar g ^*$ be the minimum $\bar g$ over all POVMs, and let
\[ b^* = b(\pi) \braket{s_0,s_1}. \]

\begin{thm} \label{thm:pure}
For states $\ket{s_0}$ and $\ket{s_1}$ and prior $\pi$,
\begin{enumerate}
    \item If $g$ is convex-admissible then $\bar g^* = g(p)$ where $p$ satisfies $b(p) = b^*$, and it is achieved by a projection with $p_0 + p_1 = 1$.
    \item If $g$ is concave-admissible and the problem is not too skewed then $\bar g^* = 2 b^* g(1/2)$, and it is achieved by a three-element measurement with $p_0=0$, $p_1=1$ and $p_2=1/2$.
\end{enumerate}
\end{thm}

The proof hinges on the following.
\begin{lemma} \label{lem:r_pure}
Let $g(\cdot)$ be the Bhattacharya function $b(\cdot)$. Then, $\bar b ^* = b^*$, and it is achieved by any eligible measurement.
\end{lemma}

\begin{proof}
For this particular objective,
\begin{align*} \bar g 
&= \sum_{j=0}^{J-1} q_j b(p_j)\\
&= \sum_{j=0}^{J-1} q_j \sqrt{p_j(1-p_j)}\\
&=b(\pi)\sum_{j=0}^{J-1}\sqrt{q_{j,0}q_{j,1}},\end{align*}
where in the last equality we used \eqref{eq:Bayes}, assuming that all $q_j>0$ (otherwise just drop the operator from the set). Now, for any element,
\begin{align*} \sqrt{q_{j,0}q_{j,1}} &= \sqrt{\braket{s_0|E_j|s_0} \braket{s_1|E_j|s_1} } \\
&\geq \sqrt{|\braket{s_0|E_j|s_1}|^2}  \\
&= \left| \braket{s_0|E_j|s_1} \right|. \end{align*} 
Here, the first inequality is Cauchy-Schwartz, which holds with equality for eligible measurements. We conclude that 
\begin{align*} \bar g &\geq b(\pi) \sum_{j=0}^{J-1}  |\braket{s_0|E_j|s_1}| \\
&\geq b(\pi) \left|\sum_{j=0}^{J-1}  \braket{s_0|E_j|s_1}\right| \\
&=  b(\pi) \braket{s_0|s_1} \\ &= b^*, \end{align*}
with equality for eligible measurements, as required.
\end{proof}

\begin{proof}[Proof of Theorem~\ref{thm:pure}]
Proposition~\ref{prop:Jensen_admissible} gives lower bounds on $\bar g$ in terms of $E[b(p)]$ for both the convex and concave cases, while Lemma~\ref{lem:r_pure} gives a lower bound on $E[b(p)]$. Combining the bounds yields the required inequalities. For achievability, by Proposition~\ref{prop:exists_pure}, there exist measurements which achieve equality in both Proposition~\ref{prop:Jensen_admissible} and Lemma~\ref{lem:r_pure}. 
\end{proof}

\section{Coherent States: Dolinar-Type Receivers}

Now we consider a specific kind of pure states known as coherent states, which is of great importance in optics, and in optical communications in particular. With some abuse of notation, we say that a state $\ket{s}$ is defined by a complex waveform $\{s(t)\}$, $t\in[0,T]$. The following list summarizes the properties of coherent states that are relevant to our analysis.
\begin{enumerate}
    \item  The inner product between two states $\ket{s_0}$ and $\ket {s_1}$ with waveforms $s_0(t)$ and $s_1(t)$ is given by \eqref{eq:coherent_inner}.
    \item When two coherent states with waveforms $s(t)$ and $\ell(t)$ are used as inputs of a \emph{beamsplitter}, we have an output that is a coherent state with waveform $s(t)+\ell(t)$.
    \item When a coherent state with waveform $s(t)$ is measured by a \emph{photon detector}, the output is a Poisson process with instantaneous rate $\lambda(t)=|s(t)|^2$.
\end{enumerate}

\begin{figure}[tbp]
	\centering
	\includegraphics[width=15cm]{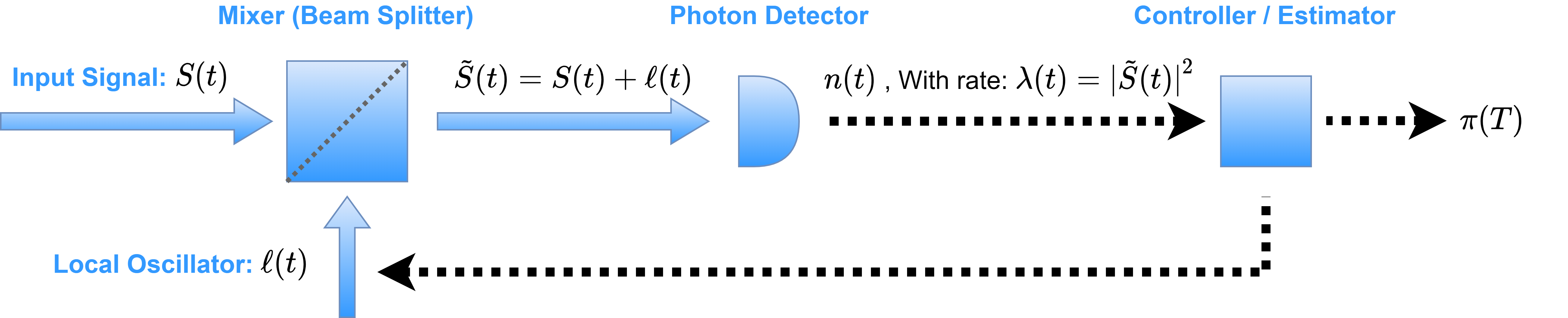}
	\caption{Dolinar receiver structure. Thick blue arrows denote optical signals.}
	\label{fig:DolinarReceiver}
\end{figure}

A Dolinar-type receiver is comprised of: (see Figure~\ref{fig:DolinarReceiver})

\begin{itemize}
	\item A photon detector, which produces a counting process $n(t)$
    \item A local oscillator, which emits a complex signal $\ell(t)$. This signal may depend upon 
    \[ n^t \triangleq \{n(t'): \ t' \in [0,t)\} .\]
    in an arbitrary manner.
    \item A beamsplitter, which creates $\tilde S(t) = S(t) + \ell(t)$.
    \item A posterior probability estimator, which computes for any time $t$ 
    \begin{IEEEeqnarray}{r.C.l} 
        \pi(t) &=& \Pr(X=1|n^t). \label{eq:pi_T}
    \end{IEEEeqnarray}
\end{itemize}

Identifying $\pi(T)$ as the measurement outcome posterior $p$, this receiver is an instance of a measurement for discrimination between the coherent states. For example, if we declare $\hat X = 0$ when $p(T)>1/2$ and $\hat X = 1$ otherwise, we have discrimination in the error-probability sense. Indeed, Dolinar has found that with a judicious choice of dependence of the local signal on $n^t$, one  may achieve the Helstrom bound \eqref{eq:Helstrom}. 
The signal design is a special case of a sequential decision problem, an in particular of sequential experimental design, see, e.g., \cite{burtini2015survey} and the references therein. Such problems are in general very complex, and closed-form optimal solutions are scarce. Often, the general solution depends on the time-horizon, that is, the choice of experiment at time $t$ depends on the decision time $T>t$. However, Dolinar's solution possesses a few remarkable properties, somewhat reminiscent of optimal causal filtering in the Gaussian setting (the Kalman filter), where the experiment is a linear measurement that need not be designed:
\begin{enumerate}
	\item The solution is independent of the time horizon. That is, the optimal $\ell(t)$ produces a posterior $\pi(t)$ which gives optimal error probability also at any time $t<T$. In other words, there is no ``exploration-exploitation trade-off''.
	\item The performance is independent of the measurement outcome process. That is, if we use the optimal $\ell(t)$, the error probability given the Poisson process $\{n(t)\}$ is always the same.
	\item The optimal $\ell(t)$ is one of two pre-selected functions, where the photon arrival process merely determines the choice between them.
	\item The solution also maximizes the mutual information between the state and the measurement outcome (\cite{Lizhong2}, see also \cite{DolinarNew}).
\end{enumerate}

Indeed the optimality and elegance of Dolinar's receiver have motivated many works. In \cite{Geremia_2004}, various receiver strategys are compared and imperfect detection is considered. Real-time calibration is considered in \cite{Bilkis_2020,Bilkis2021ReinforcementlearningCO}, where the latter applies reinforcement learning. Experimental evidence of receiver implementation is provided in \cite{experimental_1,experimental_2}. Extensions beyond a binary constellation are provided in \cite{Becerra2011MarystatePD,Rosati_2017_2,Muller_2015,Rosati_2021,Ferdinand_2017}. In \cite{Assalini_2011}, a connection to multiple-copy state discrimination is made. This connection is very natural, and we comment about this line of work at the end of this section.  

We explain the remarkable properties of the receiver, by saying that for any $t$ (with Dolinar's choice of local signal) it materializes the backward channel of Figure~\ref{fig:channel1}. Since in this channel the error probability is independent of the measurement outcome, the evolution of $p(t)$ is ``almost deterministic'', that is, the only randomness is in the choice between $(p,1-p)$, which translates to the choice between the two pre-selected functions. Indeed if this is the case, then following the same considerations the same solution is optimal for any relatively convex objective. We prove this, and also provide the optimal strategy in the relatively concave case.

We now define the optimal strategies. We assume without loss of generality that $s_1(t)\geq s_0(t)$ are real (this can always be obtained by information-preserving pre-processing of adding a fixed local signal and phase-shifting), and that the waveforms are finite and continuous. We use a discretization of time, that is, at time $t=k\tau$ we set $\ell(t)=\ell[k]$ and keep it fixed for duration $\tau$. We measure $n[k] = n(k\tau)$, where for multiple arrivals within the same epoch we only count one, and $\ell[k]$ will be a function of $n^k \triangleq \{n[k], k=0,\ldots,k\}$, $s_0[k] = s_0(k \tau)$ and $s_1[k] = s_1(k \tau)$. We also keep track of the posterior distribution, $p[k] = \Pr(X=1|n^k)$, where we initialize $p[0]=\pi$. The optimal strategy is obtained in the limit $\tau \rightarrow 0$. By continuity of the waveforms, the limit exists.\footnote{the continuity assumption is not necessary, and the results can probably be extended to any measurable functions of time; however, it allows us to avoid mathematical subtleties.} Let 
\[ \Delta[k] = \sum_{j=1}^k (s_1[k]-s_0[k])^2 . \] 

\begin{enumerate}
    \item Convex-optimal strategy. Let $p[k]$ satisfy
    \[ b(p[k]) = b(\pi) \exp\{-\Delta[k]\}, \]
    where if the sum of $n^k$ is even (resp. odd) we take the smaller (resp. larger) root. We use the Dolinar signal $\ell[k] = \ell_D(s_0[k],s_1[k],p[k])$, where
    \begin{align} \label{eq:Dolinar_signal}  \ell_D(s_0,s_1,\pi) = \frac{s_0 \pi - s_1(1-\pi)}{1-2\pi}. \end{align}
    Notice that $\ell[k]$ is undefined at $k=0$ if $\pi = 1/2$; in that case, assume $\pi = 1/2 - \delta$ where $\delta$ is a small positive number.
     \item Concave-optimal strategy. Here we have two stages.
     \begin{enumerate}
         \item As long as 
         \[ \Delta[k] < \log \frac{\pi}{1-\pi}\]
         use $\ell[k] = - s_1[k]$ (notice that this never happens for $\pi = 1/2$). If a photon arrives, halt with $\pi[k] = 0$.
         \item Otherwise, alternate between $\ell[k] = - s_0[k]$ and $\ell[k] = - s_1[k]$. If a photon arrived with the former (resp. latter), halt with $\pi[k]=1$ (resp. $\pi[k]=0$).
     \end{enumerate}
\end{enumerate}
Notice that both strategies are independent of the time horizon.

For an objective $g(\cdot)$ let $\bar g(T)$ be the expected mean, optimized over all possible strategies for the Dolinar structure. Let
\[ b^*(T) = b(\pi) \exp\{-\Delta(T)\}. \]
Our main result is as follows.
\begin{thm} \label{thm:Dolinar}
For signals $s_0(t)$ and $s_1(t)$ and prior $\pi$,
\begin{enumerate}
    \item If $g$ is convex-admissible then $\bar g^*(T) = g(p)$ where $p$ satisfies $b(p) = b^*(T)$, and it is achieved by the convex-optimal strategy above in the limit $\tau\rightarrow 0$ (and $\delta\rightarrow 0$ if applicable)
    \item If $g$ is concave-admissible and 
    \[  \Delta(T) < \log \frac{\pi}{1-\pi}, \]
        then $\bar g^* = 2 b^* g(1/2)$, and it is achieved by the concave-optimal strategy above in the limit $\tau\rightarrow 0$.
\end{enumerate}
\end{thm}

The converse part of the theorem (not only for the Dolinar structure but for any detector) follows from the converse part of Theorem~\ref{thm:pure}, with the inner product \eqref{eq:coherent_inner}. However, we prefer to prove it directly from the convexity results, such that the proof does not use any `quantum'' argument.

We make a ``local'' argument, for short durations, where each such duration will be identified later with a single epoch of the strategies above. 
\begin{lemma}\label{lem:local}
Let $X$ be binary with $\Pr\{X=1\}=\pi\leq 1/2$, let $Y|X=i$ be Poisson with parameter $\tau \lambda_i$, where $\lambda_i = |s_i+\ell|^2$ for some real $s_1 \geq s_0$, and let $p_j=\Pr\{X=1|Y=j\}$ be the posterior. Let
\[ \delta b = \lim_{\tau\rightarrow 0} \frac{b(\pi) - E[b(p)]}{\tau} \]
be the infinitesimal improvement in the Bhattacharyya parameter \eqref{eq:Bhattacharyya}. 
Then:
\begin{enumerate}
    \item $\delta b \leq b(\pi) |s_1-s_0|^2$, with equality for all $\ell \notin (-s_0,-s_1)$.
    \item Specifically for the Dolinar signal $\ell=\ell_D(s_0,s_1,\pi)$ \eqref{eq:Dolinar_signal}, $p_1=1-\pi$.
\end{enumerate}
\end{lemma}

The proof is given in Appendix~\ref{app:local}. We are now in position to prove the global result.

\begin{IEEEproof}[Proof of Theorem\ref{thm:Dolinar}]
Identifying $\pi$, $s_0$ and $s_1$ of Lemma~\ref{lem:local} with $p[k]$, $s_0[k]$ and $s_1[k]$, respectively, the bound means that for any strategy,
\[ \frac{db(\pi(t))}{dt} \geq - (s_1(t)-s_0(t))^2 b(\pi(t)), \]
which immediately yields:
\begin{align} \label{eq:Dolinar_ineq} b(p(T)) \geq b(\pi) \exp \{-\Delta(T)\}. \end{align}

\emph{Convex case}: The inequality stems from \eqref{eq:Dolinar_ineq} together with the first part of Proposition~\ref{prop:Jensen_admissible}. To see that it is achievable by Dolinar's signal, notice that this signal satisfies equality in Lemmma~\ref{lem:local} and consequently in \eqref{eq:Dolinar_ineq}, thus it is left to show that with this signal $b(p(T))$ is deterministic, yielding equality in Proposition~\ref{prop:Jensen_admissible}. To see this, notice that by the second part of Lemma~\ref{lem:local}, if a photon arrived $b$ remains fixed.

\emph{Concave case}: The inequality stems from \eqref{eq:Dolinar_ineq} together with the second part of Proposition~\ref{prop:Jensen_admissible}. To see that it is achievable by the concave strategy, notice that this signal satisfies equality in Lemmma~\ref{lem:local} and consequently in \eqref{eq:Dolinar_ineq}, thus it is left to show that with this strategy, $p(T) \in \{0,1/2,1\}$. Trivially, if a photon arrived when $\ell \in \{-s_0,-s_1\}$ the posterior will be certain. Direct calculation shows, that if $T$ is large enough according to the Theorem condition, we reach $\pi(T) = 1/2$.
\end{IEEEproof}

We can easily identify the connections between the details of these optimal solutions, and those of the optimal solutions for general pure states. Dolinar's solution dictates that whenever a photon arrives, the posterior exactly flips, thus for any horizon $T$ the whole measurement outcome process is equivalent to a projection on the states yielding a symmetric posterior. For the concave case, when the prior is uniform, the fast flipping between $-s_0(t)$ and $-s_1(t)$ keeps it symmetric until a photon arrives, thus it materializes a three-element symmetric POVM. When the prior is not uniform, an initial stage precedes, where we only try to rule out the less probable option, until if no photon arrived the posterior is uniform. The first and second stages together materialize the measurement; if the energy of the signals difference until the horizon is large enough with respect to the prior, then these two stages together achieve the optimal three-element measurement.

We conclude by pointing out the connection with multiple-copy state discrimination theory. Indeed, we approximated the waveforms as piecewise-continuous. Equivalently, we had to distinguish between two sequences of constant-amplitude coherent pulses, which is a special case of sequences of pure states. This general class of problems was considered in, e.g., \cite{Ac_n_2005,Brandsen2019Adaptive_2,Brandsen2020ReinforcementLW}, with an emphasis on optimal strategies that apply ``local'' measurements to the individual states. Specifically, Theorem 2 of \cite{Brandsen2019Adaptive_2} directly generalizes the properties of the Dolinar receiver to such sequences of states. Indeed, using the approach of the current work, one may easily extend this result to convex-admissible and concave-admissible objectives, and then obtain our Theorem~\ref{thm:Dolinar} as a special case. However, notice that in this section we only used ``classical'' analysis, that is, the Poisson measurement statistics.

\section{Conclusion}

We have derived the optimal measurement between pure states for a variety of criteria, using only simple considerations such as the Cauchy-Schwartz and convexity (Jensen) inequalities. We obtain well-known optimal measurements as special cases. For an optical detector utilizing photon counting and feedback, we obtain these results using the same convexity considerations, without resorting to any quantum considerations. By proving a relative convexity relation between Renyi entropies, We have shown that ``most'' of these entropies qualify as measures that fall within our framework.

It is natural to ask whether these results extend beyond the cases we considered. Such extensions could be objective functions not related to the Bhattacharyya distance, more than two states, or non-pure states. However, the simple methods we have used do not seem to readily apply to any of these.

\begin{appendices}

\section{Backward Channels}
\label{app:backward}

In this section we consider the well-known solutions to the problems of minimum error probability and unambiguous discrimination. We demonstrate that the well-known solutions are compatible with the extremal channels of Figure~\ref{fig:channels}, and have the performance guaranteed by Proposition~\ref{prop:exists_pure}.

In both of them, it turns out that the optimal measurement operators always remain in the subspace spanned by the states. Thus we may think of a two-dimensional Hilbert space, and further we may w.l.o.g. consider real-valued vectors. Let the angle between these two vectors be $\theta$, where $\cos(\theta) = \braket{s_0|s_1}$.

We use the notation defined in Section~\ref{sec:pure}: the prior is $\Pr(X=1)=\pi \leq 1/2$, the transition probabilities are $\Pr(\hat X = j|X=i)=q_{i,j}$, the outcome probabilities are $\Pr(\hat X = j) = q_j$ and the posteriors are $\Pr(X=1|\hat X = j)= q_j$; these quantities are related by Bayes' rule \eqref{eq:Bayes}.

\emph{Error probability}. It is well known that the optimal measurement is a projection. Denote the angle between the bisector of the projection and that of the states by $\phi/2$. We have that
\begin{align*} q_{0,1} & = \cos^2 \left( \frac{1}{2} \left(\frac{\pi}{2}+\theta+\phi\right)\right)\\
q_{1,0} & = \cos^2 \left( \frac{1}{2} \left(\frac{\pi}{2}+\theta-\phi\right)\right). \end{align*} Optimizing the error probability $p_e = (1-\pi) q_{0,1} + \pi q_{1,0}$ w.r.t. $\phi$ gives the condition
\begin{align} \label{eq:pe_cond} (1-\pi) \cos(\theta+\phi) = \pi \cos (\theta-\phi). \end{align}
Now, we claim that the optimal backward channel satisfies that the error event is independent of the measurement outcome, that is, $p_1 = 1-p_0$. Applying Bayes' rule, we need to show that
\begin{align} \label{eq:pe_need} \frac{q_{0,1}q_0}{q_{1,0}q_1} = \frac{\pi}{1-\pi}. \end{align}
Using total probability to evaluate $\{q_i\}$ and applying basic manipulations, we have that
\[ \frac{q_{0,1}q_0}{q_{1,0}q_1} = \frac{\pi A(\theta,\phi) + (1-\pi) \cos^2(\theta+\phi)}{(1-\pi)  A(\theta,\phi) + \pi \cos^2(\theta-\phi)} , \]
for $A(\theta,\phi) = (1+\sin(\theta+\phi))(1+\sin(\theta-\phi))$.
Substituting \eqref{eq:pe_cond} in this expression indeed shows \eqref{eq:pe_need}. It is easy to see that the achieved error probability satisfies the first part of Proposition~\ref{prop:exists_pure}, and in particular the Helstrom bound \eqref{eq:Helstrom}. 
Interestingly, this channel is exactly the rate-distortion function (RDF) achieving test channel for a Bernoulli-$\pi$ variable under the Hamming distortion measure.

    \emph{Unambiguous discrimination}. Obviously, certainty is possible only when the operator is orthogonal to one of the states. In a two-dimensional space that means that we have rank-1 operators $a_i \ket{\psi_i}\bra{\psi_i}$, $i \in \{0,1\}$, where the coefficients are non-negative, and $\braket{s_0|\psi_1} = \braket{s_0|\psi_0} = 0$.
    These two operators will yield the certain results, while the third will lead to uncertainty. Since the third operator must be non-negative, we require the larger eigenvalue of $\alpha_0\ket{\psi_0}\bra{\psi_0} + \alpha_1\ket{\psi_1}\bra{\psi_1}$ to be at most $1$. Evaluating explicitly, this leads to the conditions (for positive coefficients):
    \begin{subequations}
    \begin{align}
        \left(1-\frac{1}{\alpha_0}\right)\left(1-\frac{1}{\alpha_1}\right) &\leq \cos^2(\theta) \label{eq:as_constraint1} \\
        \alpha_0,\alpha_1 & \leq 1. \label{eq:as_constraint2}
    \end{align}
    \label{eq:as_constraints}
    \end{subequations}
    Under these constraints, we wish to choose coefficients that maximize the probability of certainty, that is,
    \begin{align*} \pi q_{1,1} + (1-\pi) q_{0,0} &= \pi \alpha_1 \braket{s_1|\psi_1} + (1-\pi) \alpha_0 \braket{s_0|\psi_0} \\ &= (\pi \alpha_1 + (1-\pi) \alpha_0) \sin^2(\theta). \end{align*}
    Thus, we need to maximize $\pi \alpha_1 + (1-\pi) \alpha_0$ over the coefficients satisfying \eqref{eq:as_constraints}. Direct calculations show that the optimizer under \eqref{eq:as_constraint1} alone is given by:
    \begin{subequations}
    \begin{align}
        a_0 &= \frac{1-\sqrt\frac{1-\pi}{\pi} \cos(\theta)} {\sin^2(\theta)} \\ 
                a_1 &= \frac{1-\sqrt\frac{\pi}{1-\pi} \cos(\theta)} {\sin^2(\theta)}.
    \end{align}
    \label{eq:as_opt}
    \end{subequations}

    We now have two cases:
    \begin{enumerate}
      \item Not too skewed case: $\pi/(1-\pi) \geq \cos^2(\theta)$. In this case the solution \eqref{eq:as_opt} gives non-negative coefficients satisfying \eqref{eq:as_constraint2} as well, thus it is the optimizer. Since the eigenvalue constraint is met with equality, the third operator must be rank-$1$ as well. It can be verified that the third measurement vector ``scans'' from the bisector of states at $\pi=1/2$ to $\ket{s_1}$ at the limit of allowed skewness, thus indeed we have an eligible measurement. It is easy to see that the probability for uncertainty is $2\sqrt{\pi(1-\pi)}\cos(\theta)$, which shows the second part of Proposition~\ref{prop:exists_pure}. Also, direct calculation shows that with this solution 
      \[ \frac{q_{0,E}}{q_{1,E}} = \frac{1-\pi}{\pi}, \]
      thus the backward channel is that depicted in Figure~\ref{fig:channel2}. Interestingly, this channel is exactly the RDF achieving test channel for a Bernoulli variable under the erasure distortion measure, when it is ``not too skewed'', in that case $\pi \geq D/2$.

      \item Very skewed case: Otherwise, we have that in \eqref{eq:as_opt} one of the coefficients is negative and the other greater than one. It turns out that in that case the optimal solution is $\alpha_0=0$ and $\alpha_1=1$, that is, the measurement is projection on the basis of $\ket{s_0}$.
            
                   \end{enumerate}

\section{Relative Convexity of R\'enyi Entropies} \label{app:Renyi}

In this appendix we prove Theorem~\ref{thm:Renyi}. First notice that for any order $\alpha$, 
\[ h_0(p) = \left\{ \begin{array} {c c} 0 & p\in\{0,1\} \\ 1 & \text{otherwise.} \end{array} \right. \]
As this is a concave function, $h_0(\cdot) \prec h_\alpha(\cdot)$. It is also easy to see that for all $\alpha$, $h_\alpha(\cdot) \prec h_\infty(\cdot)$. As the case $\alpha=1$ holds by continuity, it is enough to consider regular orders for which \eqref{eq:Renyi} reduces to
\[ h_\alpha(p) = \frac{1}{1-\alpha} log \left(p^\alpha + (1-p)^\alpha \right) . \] As the functions are smooth, we can use the criterion \eqref{eq:derivatives_ratio}. That is, we need to show:
\begin{IEEEeqnarray}{r?C?l} 
	 \frac{h_\alpha''(p)}{h_\alpha'(p)} \geq \frac{h_\beta''(p)}{h_\beta'(p)}  \ \forall \alpha \geq \beta > 0,  \ 0<p<1/2. \label{eq:Renyi_derivatives}
\end{IEEEeqnarray}

We define:
\begin{align} \label{eq:Pk} P_k(\alpha) = k(\alpha-1)(2\alpha)^{k-1} +k\alpha(2\alpha-1)^{k-1} - (2\alpha)^k - (2\alpha-1)^k + (\alpha+1)^k - (\alpha-1)^k - 2k\alpha^{k-1} + k\alpha + 1.
\end{align}
We present two lemmas, which together show the desired result, and then prove them.

\begin{lemma} \label{lem:Renyi_1}
The condition \eqref{eq:Renyi_derivatives} holds if and only if
\[ \sum_{k=3}^\infty \frac{P_k(\alpha)}{k!} t^k \geq 0 \ \ \forall \alpha>0 . \]
\end{lemma}

\begin{lemma} \label{lem:Renyi_2}
For all integer $k \geq 3$ and for all $\alpha>0$, $P_k(\alpha) \geq 0$
\end{lemma}

\begin{proof}[Proof of Lemma~\ref{lem:Renyi_1}]
The condition \eqref{eq:Renyi_derivatives} is equivalent to the condition that 
\[ \frac{h_\alpha'(p)}{h_\beta'(p)} = \frac{1-\beta}{1-\alpha} \cdot \frac{\((1-p)^{\alpha-1} - p^{\alpha-1}\) \cdot \((1-p)^{\beta} + p^{\beta}\)}{\((1-p)^{\beta-1} - p^{\beta-1}\) \cdot \((1-p)^{\alpha} + p^{\alpha}\)}
 \] is non-decreasing in $p$. 
We make a change of variable, letting \[ t = \log \frac{1-p}{p}. \] Note that $t$ is a {\it decreasing} function of $p$. Substituting $p = \frac{1}{\exp(t)+1}$ and $1-p = \frac{\exp(t)}{\exp(t)+1}$ and simplifying, an equivalent condition is that
$F(\alpha,t) / F(\beta,t)$ is non-increasing in $t$ for all $\alpha \ge \beta \ge 0$ and $t \ge 0$, where
\[ F(\alpha,t) = \frac{\exp\((\alpha-1)t\) - 1}{\exp\((\alpha t\) + 1}. \]
Taking the derivative of $F(\alpha,t) / F(\beta,t)$, an equivalent condition is that
\[ \frac{\partial}{\partial t} F(\alpha,t) F(\beta,t) - \frac{\partial}{\partial t} F(\beta,t) F(\alpha,t) \leq 0. \]
Defining 
\[ G(\alpha,t) = \frac{\frac{\partial}{\partial t} F(\alpha,t)}{F(\alpha,t)}, \] an equivalent condition is that $G(\alpha,t)$ is non-increasing in $\alpha$.
Computing this function explicitly, we have that $G(\alpha,t) = U(\alpha,t)/D(\alpha,t)$, where
\begin{align*}
U(\alpha,t) & = -\exp \left( (2\alpha-1)t\right) + \alpha \exp\left(\alpha t\right) + (\alpha-1)\exp\left( (\alpha-1)t\right) \\
D(\alpha,t) & = \left( \exp\left( (\alpha-1)t\right) - 1\right) \cdot \left(\exp\left(\alpha t\right) + 1\right) \end{align*}
Taking the derivative of $G(\alpha,t)$ \emph{now with respect to $\alpha$}, an equivalent condition is that
\[ \frac{\partial U(\alpha,t)}{\partial \alpha} \cdot D(\alpha,t) - \frac{\partial D(\alpha,t)}{\partial \alpha} \cdot U(\alpha,t) ~\le~ 0.
\]

\noi Computing explicitly and simplifying, the condition becomes that for all $\alpha, t \ge 0$ holds that $P(\alpha,t) \geq 0$, where
\begin{align*} \Phi_\alpha(t) &= 
\Big((\alpha-1)t - 1\Big) \exp(2 \alpha t) + \Big(\alpha t - 1\Big) \exp\left((2\alpha-1)t\right) + \exp\left((\alpha+1)t\right) - 2t\exp(\alpha t) - \exp\left((\alpha-1)t\right) \\ & \ \ \ + \Big(\alpha t+1\Big) \exp(t) + \Big((\alpha-1) t + 1\Big).
\end{align*}
Finally, we compute the Taylor expansion of $P_\alpha(t)$ around $t=0$, that is, 
\[ \Phi_\alpha(t) = \sum_{k=0}^\infty \frac{P_k(\alpha)}{k!} t^k . \]
In order to find the coefficients, substitute in the expression for $\Phi_\alpha(t)$
\begin{align*} \exp (\gamma t) &= \sum_{k=0}^\infty \frac{\gamma^k}{k!} t^k \\
t \exp (\gamma t) &= \sum_{k=1}^\infty \frac{\gamma^{k-1}}{(k-1)!} t^k \end{align*}
where $\gamma$ represents the appropriate coefficient of $t$ in the exponent. 
One may verify that $P_k(\alpha) \equiv 0$ for $k = 0,1,2$, while for $k\geq 3$ $P_k(\alpha)$ is indeed given by \eqref{eq:Pk}.
\end{proof}

\begin{proof}[Proof of Lemma~\ref{lem:Renyi_2}]
Consider $P_k(\cdot)$ \eqref{eq:Pk}. It is easy to verify that $P_k(1)=0$.\footnote{In fact, this is a root of order $2$, but we did not find a way to use that.} We can thus perform division to find the polynomial $Q_k(\alpha)$ satisfying $P_k(\alpha) = (\alpha-1) Q_k(\alpha)$. To that end, we rearrange $P_k(\alpha)$ \eqref{eq:Pk} as follows.
\begin{align*}
    P_k(\alpha) &= k(\alpha-1) (\alpha^{k-1} + (2\alpha)^{k-1}) - (\alpha-1)^k \\
    &- ((2\alpha-1)^k - 1)  \\
    &- ((2\alpha)^k - (\alpha+1)^k) \\
    &+ k \alpha ((2\alpha-1)^{k-1} - \alpha^{k-1}) \\
    &- k \alpha (\alpha^{k-2} - 1).
    \end{align*}
Now we treat each line separately. After the first trivial one, we apply the identity
\[ b^r - a^r = (b-a) \sum_{i=0}^{r-1} a^i b^{r-1-i} \] to the others. Putting all together and rearranging, we find that:
\[ Q_k(\alpha) = \tilde Q_k(\alpha) + R_k(\alpha), \]
where
\begin{align} \label{eq:tQk} \tilde Q_k(\alpha) =  
k (2\alpha)^{k-1} - (\alpha-1)^{k-1} -
\sum_{i=0}^{k-1} (2\alpha)^i(\alpha+1)^{k-1-i} +
2  k \alpha^{k-1} + 
2 \sum_{i=0}^{k-1}  (2\alpha - 1)^i
\end{align}
and 
\begin{align*}
   R_k(\alpha) = k\alpha \left( \sum_{i=0}^{k-3} \alpha^i (2\alpha-1)^{k-2-i} - \sum_{i=0}^{k-3} \alpha^i \right).
\end{align*}
It suffices to show that both these terms are non-negative for $\alpha>1$ and non-positive for $0<\alpha<1$. For $R_k(\alpha)$ it is trivial, thus the rest of the proof is dedicated to showing the same property for $\tilde Q_k(\alpha)$ as well (treating the two cases separately). 

\underline{The case $\alpha>1$}. We rearrange the polynomial \eqref{eq:tQk} as follows. 
\begin{align*} 
\tilde Q_k(\alpha) &=  (k-1) (2\alpha)^{k-2} (\alpha+1) -  \sum_{i=0}^{k-2} (2\alpha)^i(\alpha+1)^{k-1-i} \\
&  + (k-3) \Bigl[ \alpha (2\alpha-1)^{k-2} + \alpha^{k-1} + \Big((2\alpha)^{k-1} - (2\alpha)^{k-2}(\alpha+1)\Big) - 2 \cdot (2\alpha - 1)^{k-2}   \Bigr] \tag{a} \\
& + 2 \left( (k-3)(2\alpha-1)^{k-2} - \sum_{i=2}^{k-2}  (2\alpha - 1)^i   \right) \\
&+ 2 \left( \alpha (2\alpha-1)^{k-2} + \Big((2\alpha)^{k-1} - (2\alpha)^{k-2}(\alpha+1)\Big) - (2\alpha - 1)^{k-1} \right) \tag{b} \\
&+ \alpha (2\alpha-1)^{k-2} + \alpha^{k-1} - 2 \cdot (2\alpha-1) \tag{c} \\
&+ 2(\alpha^{k-1}-1) - (\alpha-1)^{k-1}. \tag{d}
\end{align*}
We claim that for any $\alpha>1$ and $k\geq 3$, all lines are non-negative. We provide a proof for the non-trivial expressions, tagged (a)-(d). 
To see that (a) is non-negative, note that it suffices to prove that $(2\alpha)^{k-1} - (2\alpha)^{k-2}(\alpha+1) \ge (2\alpha - 1)^{k-2} - \alpha^{k-2}$. Dividing out by $\alpha-1$, it suffices to prove that
\[
(2\alpha)^{k-2} ~\ge~ (2\alpha-1)^{k-3} + \alpha (2\alpha-1)^{k-4} +...~ \alpha^{k-3}
\]
for all $k \ge 3$ and $\alpha \ge 1$. It is easy to see that this holds for $k=3$. We proceed by induction on $k$. Let $L_k$ and $R_k$ denote that LHS and the RHS of the above. We have that
\begin{align*}
L_{k+1} - R_{k+1} &= 2\alpha \cdot L_k - \Big((2\alpha-1) \cdot R_k + \alpha^{k-2}\Big) \\ &\ge L_k - \alpha^{k-2} \\ &> 0.
\end{align*}
The non-negativity of (b) is equivalent to \[ (\alpha-1)(2\alpha)^{k-2} \ge (\alpha-1)(2\alpha-1)^{k-2}. \]
The non-negativity of (c) follows from \begin{align*} \alpha (2\alpha-1)^{k-2} & \ge \alpha^{k-1} \\ &= (1 + (\alpha-1))^{k-1} \ge 1 + (k-1) \cdot (\alpha-1) \\ & \ge 2 \alpha-1. \end{align*}
Finally for (d), we have
\begin{align*}
 \alpha^{k-1}-1 - \frac{(\alpha-1)^{k-1}}{2}   &\geq
 \alpha^{k-1}-1 - (\alpha-1)^{k-1} \\ &=
 \left( 1 + (\alpha-1)\right) ^{k-1} - \left(1 + (\alpha-1)^{k-1}\right) \\
 &\geq 0.
\end{align*}

\underline{The case $0<\alpha<1$}. We rearrange the polynomial \eqref{eq:tQk} in a different way.
\begin{align*} 
-\tilde Q_k(\alpha) &= 
(k-1) (2\alpha)^{k-1} -
\sum_{i=1}^{k-1} (2\alpha)^i(\alpha+1)^{k-1-i} \\
&+  (\alpha+1)^{k-1} + (\alpha-1)^{k-1} - (2\alpha)^{k-1} \tag{a} \\
&+ 2\left( \sum_{i=0}^{k-1} (2\alpha - 1)^i - k\alpha^{k-1}    \right). \tag{b}
\end{align*}
Again we claim that all lines are non-negative (this time for $\alpha<1$) and provide a proof for the non-trivial expressions, tagged (a) and (b). 
For (a), rewrite the non-negativity claim as \[ \left(\frac{\alpha+1}{2\alpha}\right)^{k-1} + \left(\frac{\alpha-1}{2\alpha}\right)^{k-1} \ge 1. \] Set $y = \frac{1-\alpha}{2\alpha}$. Note that $y \ge 0$ and the claim above becomes $(1+y)^{k-1} \ge 1 - (-y)^{k-1}$, which follows from a stronger and immediate claim \[ (1+y)^{k-1} \ge 1 + y^{k-1}. \] For (b), rewrite the non-negativity claim as \[ \frac{(2\alpha-1)^k - 1}{\alpha-1} \ge 2k\alpha^{k-1}. \] Rearranging, this is the same as
\[
2k (1-\alpha) \alpha^{k-1} + (2\alpha-1)^k ~\le~ 1,
\]
for all $0 \le \alpha \le 1$ and for all $k \ge 3$. We will actually show this for all $k \ge 1$. Let $\delta = 1 - \alpha$. Then $\delta \ge 0$. Write
%\begin{align*}
%(2\alpha-1)^k &= (\alpha - \delta)^k \\ &= \sum_{i=0}^k (-1)^i {k \choose i} \alpha^{k-i} \delta^i.
%\end{align*}
%Hence
\begin{align*}
2k (1-\alpha) \alpha^{k-1} + (2\alpha-1)^k &= 2k \alpha^{k-1} \delta + (\alpha-\delta)^k 
\\ &= 2k \alpha^{k-1} \delta + \sum_{i=0}^k (-1)^i {k \choose i} \alpha^{k-i} \delta^i
\\ &= \alpha^k + {k \choose 1} \alpha^{k-1} \delta + \sum_{i=2}^k (-1)^i {k \choose i} \alpha^{k-i} \delta^i \\ &\le
\sum_{i=0}^k {k \choose i} \alpha^{k-i} \delta^i \\ &= (\alpha + \delta)^k \\ &= 1.
\end{align*}

Since in these two cases we have shown that $(\alpha-1)Q_k(\alpha)\geq 0$ for all $\alpha\geq 0$ and $k\geq 3$, the proof is completed.

\end{proof}

\section{Local Analysis}
\label{app:local}

In this section we prove Lemma~\ref{lem:local}.

\begin{figure}[t]
	\centering
	\includegraphics[scale=0.5]{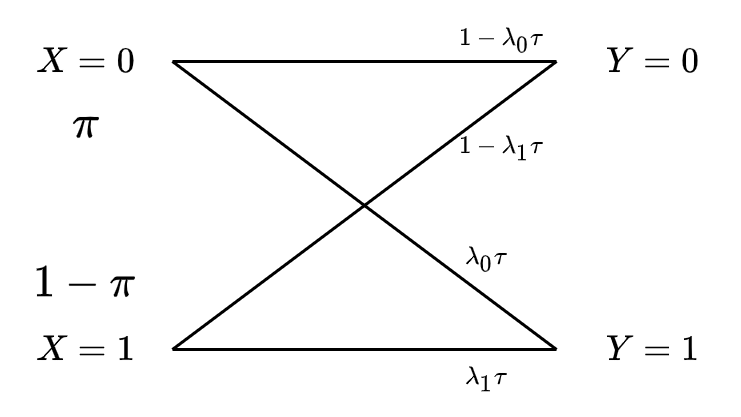}
	\caption{Corresponding binary channel in a short time interval. Note that since $\tau\rightarrow0$, with high probability there is no photon arrival.}
	\label{fig:BinaryChannel}
\end{figure}

Since $Y\sim \text{Poisson}(\lambda_i\tau)$, it holds that $\Pr(Y>1)=O(\tau^2)$. This allows us to discard the values of $Y>1$ and consider the binary channel from $X$ to $Y$ where
\[ \Pr(Y=1|X=i) = \tau\lambda_i. \]
The corresponding binary channel can be seen in Figure~\ref{fig:BinaryChannel}.
We now define further quantities for this channel. The normalized prior on $Y$ is,
\begin{IEEEeqnarray}{r/C/l} 
    \bar \lambda    &\triangleq&  \frac{\Pr(Y=1)}{\tau} \noN 
\\                  &=&           \pi \lambda_0 + (1-\pi) \lambda_1. \label{eq:lam} 
\end{IEEEeqnarray}
The posteriors can be written as:
\begin{IEEEeqnarray}{r/C/l} 
    p_1 &=& \Pr(X=0|Y=1)\noN
\\      &=& \frac{\Pr(Y=1|X=0)\Pr(X=0)}{\Pr(Y=1)}\noN
\\      &=& \frac{\pi \lambda_0}{\bar \lambda}, \label{eq:p1} 
\end{IEEEeqnarray}
and
\begin{IEEEeqnarray}{r/C/l} 
p_0 &=& \Pr(X=0|Y=0)\noN
% \\      &=& \frac{\Pr(Y=0|X=0)\Pr(X=0)}{\Pr(Y=0)}\noN
\\      &=&  \frac{1-\lambda_0 \tau}{1-\bar\lambda \tau} \pi \noN
\\      &=&  [1 - (\lambda_0 - \bar \lambda) \tau] \pi + O(\tau^2) \noN
\\      &=&  \pi - \bar \lambda \tau (p_1-\pi) + O(\tau^2). \label{eq:p0}
\end{IEEEeqnarray}
Plugging these posteriors into the definition of $\delta b$, we have:
\begin{IEEEeqnarray}{r/C/l}  
    \delta b
        &=& \lim_{\tau\rightarrow{0}}\frac{1}{\tau}\left[b(\pi)-E_Y[b(\Pr(X=0|Y=y)])\right]\noN 
\\      &=& \lim_{\tau\rightarrow{0}}\frac{1}{\tau}\left[b(\pi)-(1-\bar\lambda\tau)b(p_0)-\bar\lambda\tau b(p_1)\right] \noN
\\      &=& \lim_{\tau\rightarrow{0}}\frac{1}{\tau}\left[b(\pi)-(1-\bar\lambda\tau)b(\pi+\bar\lambda\tau(p_1-\pi))-\bar\lambda\tau b(p_1)\right] \noN
\\      &=& \bar\lambda\left[(p_1 - \pi)b'(\pi)+b(\pi)-b(p_1)\right] \noN, 
\end{IEEEeqnarray}
where 
\[ b'(\pi) = \frac{1-2\pi}{2b(\pi)} \]
is the derivative of $b$ evaluated at $\pi$.
Substituting the derivative and rearranging, we find that
\begin{align} \delta b =  \frac{ \bar\lambda}{b(\pi)}\left(\sqrt{(1-\pi)p_1} - \sqrt{\pi(1-p_1)}\right)^2
. \label{eq:dr1} \end{align}
 In order to find the optimal $\delta b$ we should maximize this expression. However, it involves $\bar \lambda$ and $p_1$, both of which depend on our design parameter $\ell$. In order 
to show the optimal $\delta b$, it is convenient to perform optimization w.r.t. $p_1$ rather than $\ell$. For this, without loss of generality, we may consider only real $\ell$ (recall that we assumed real $s_0, s_1$), since an imaginary part of $\ell$ would increase both $\{\lambda_i\}$ by the same amount, which in turn may only decreases $\delta b$.
\par
Using \eqref{eq:p1} and the rate definition $\lambda_i$, we get: % the following identities,
\begin{subequations}
\begin{IEEEeqnarray}{r/C/l}
    \frac{p_1}{\pi}     &=& \frac{(s_0+\ell)^2}{\bar\lambda}. \label{eq:roman1}
\\  \frac{1-p_1}{1-\pi}   &=& \frac{(s_1+\ell)^2}{\bar\lambda}. \label{eq:roman2}
\end{IEEEeqnarray}
Which yield:
\begin{IEEEeqnarray}{r/C/l}
    \bar\lambda\left[\frac{p_1}{\pi} - \frac{1-p_1}{1-\pi} \right] &=& (s_0-s_1)[s_0+s_1+2l]\label{eq:roman3}
\\  s_0+s_1+2l &=& \sqrt{\bar\lambda}\left(\sqrt{\frac{p_1}{\pi}}\pm\sqrt{\frac{1-p_1}{1-\pi}}\right),\>\>\>\label{eq:roman4}
\end{IEEEeqnarray}
\end{subequations}
where \eqref{eq:roman3} is calculated by subtracting \eqref{eq:roman2} from \eqref{eq:roman1}, and \eqref{eq:roman4} by adding the square roots of \eqref{eq:roman1} and \eqref{eq:roman2}.
Now, by combining \eqref{eq:roman3} and \eqref{eq:roman4} we get:
\begin{IEEEeqnarray}{r/C/l}
    \bar\lambda	&=& (s_0-s_1)^2\frac{\left(\sqrt{\frac{p_1}{\pi}}\pm\sqrt{\frac{1-p_1}{1-\pi}}\right)^2}{\left[\frac{p_1}{\pi} - \frac{1-p_1}{1-\pi} \right]^2} \noN
\\      &=& \frac{\pi(1-\pi)(s_0-s_1)^2}{\left(\sqrt{(1-\pi)p_1} \pm \sqrt{\pi(1-p_1)}\right)^2}, \label{eq:lambda_bar_proof}
\end{IEEEeqnarray}
which is in terms of $p_1$ and not of $\ell$, as we wanted. Out of the two solutions, we discard the smaller since by \eqref{eq:dr1}, $\delta b$ is maximized by the larger $\bar\lambda$. Therefore, for maximal $\delta b$ it holds that 
\begin{IEEEeqnarray}{r/C/l}
    \bar\lambda &=& \frac{b^2(\pi) (s_0-s_1)^2}{\left(\sqrt{(1-\pi)p_1} - \sqrt{\pi(1-p_1)}\right)^2}. \noN
\end{IEEEeqnarray}
Substituting in \eqref{eq:dr1}, we find that indeed the maximal value of $\delta b$ is $(s_0-s_1)^2 b(\pi)$. Equality holds whenever $\ell$ yields a negative sign in the denominator of \eqref{eq:lambda_bar_proof}. Tracking the sign back to \eqref{eq:roman1}-\eqref{eq:roman2}, we see that $(s_0+\ell)$ and $(s_1+\ell)$ should have opposite signs, that is, $\ell$ is not in $(-s_1,-s_0)$.

As for $\ell_D$, explicitly calculating $\lambda_0$ and $\bar\lambda$ and substituting in \eqref{eq:p1} yields that indeed $p_1=1-\pi$.

\end{appendices}

\section*{Acknowledgments}

The authors thank Anatoly Khina, Yury Polyanskyi, Michael Ben-Or and Henry Pfister for helpful discussions.

\bibliography{pure.bib}
\bibliographystyle{unsrt}
%--------------------------------------------------------------------
%	REFERENCE LIST
%--------------------------------------------------------------------
%--------------------------------------------------------------------
\end{document}